\documentclass[11pt]{article}

\usepackage[margin=1in]{geometry}
\usepackage{setspace}
\usepackage{times}

\usepackage{amsmath}
\usepackage{amssymb}
\usepackage{bm}

\usepackage{graphicx}
\usepackage{epsfig}
\usepackage{caption}
\usepackage{booktabs}
\usepackage{multirow}

\usepackage{pgf,tikz,pgfplots}
\usetikzlibrary{arrows,shapes,plotmarks,spy,backgrounds}
\pgfplotsset{compat=1.18}

\usepackage{algpseudocode}

\usepackage{comment}
\usepackage{xcolor}
\usepackage{url}

\usepackage{hyperref}
\hypersetup{
  colorlinks=true,
  linkcolor=blue,
  filecolor=magenta,
  urlcolor=blue,
  citecolor=blue,
}

\usepackage{microtype}

\usepackage{authblk}

\usepackage{fancyhdr}
\begin{document}

\title{Deep Lightweight Unrolled Network for\\ High Dynamic Range Modulo Imaging}

\author[ ]{Brayan Monroy}
\author[ ]{Jorge Bacca}
\affil[ ]{Department of Computer Science, Universidad Industrial de Santander}
\affil[ ]{Bucaramanga, Colombia}
\affil[ ]{\texttt{\url{https://github.com/bemc22/ModuloUnrolled}}} 

\date{} 
\maketitle

\begin{abstract}
Modulo-Imaging (MI) \textcolor{black}{offers} a promising alternative for expanding the dynamic range of images \textcolor{black}{by resetting the signal} intensity when it reaches the saturation level. Subsequently, high-dynamic range (HDR) modulo imaging requires a recovery process to obtain the HDR image. MI is a non-convex and ill-posed problem where recent recovery networks suffer in high-noise scenarios. In this work, we formulate the HDR reconstruction task as an optimization problem that incorporates a deep prior and subsequently unrolls it into an optimization-inspired deep neural network. The network employs a lightweight convolutional denoiser for fast inference with minimal computational overhead, effectively recovering intensity values while mitigating noise. Moreover, we introduce the Scaling Equivariance term that facilitates self-supervised fine-tuning, thereby enabling the model to adapt to new modulo images that fall outside the original training distribution. Extensive evaluations demonstrate the superiority of our method compared to state-of-the-art recovery algorithms in terms of performance and quality.
\end{abstract}

\vspace{0.5em}

\section{Introduction}
%
%
%
%

\begin{figure}[!t]
    \centering
    \includegraphics[width=0.7\linewidth]{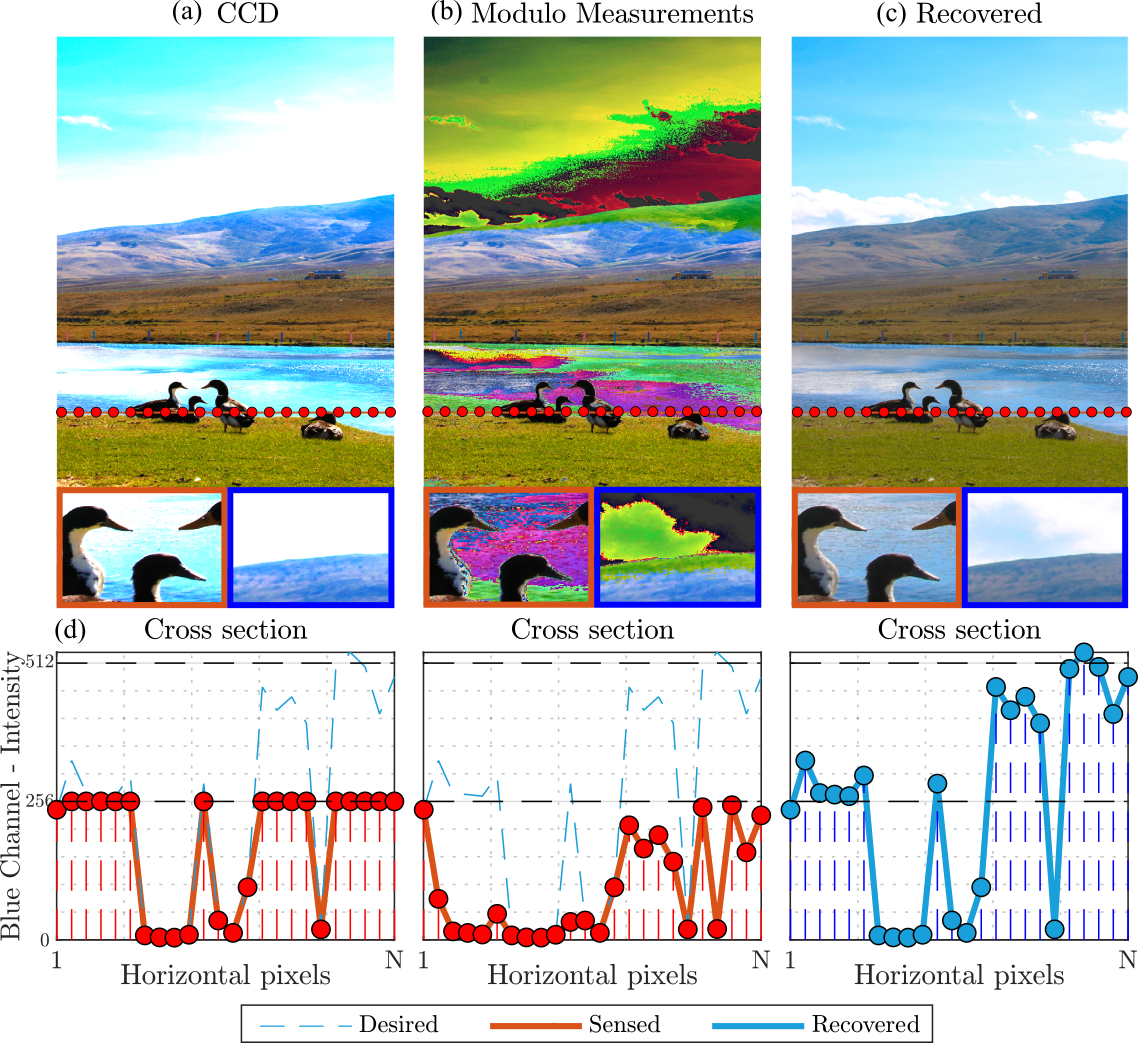} 
    \caption{ \textbf{Modulo imaging.} 
    (a) Images from conventional saturated image acquisition as CCD. (b) Images from modulo sensor that resets the pixel intensity. (c) Recovered image from the modulo image.  (d) Horizontal intensity cross-sections of each blue component image \textcolor{black}{(red dots line)} are provided to evidence the sensing and recovery process.
    \label{fig:main} } \vspace{-1em}
\end{figure}

Sensing natural images with a High Dynamic Range (HDR) presents a significant challenge for conventional imaging sensors such as CCD and CMOS, primarily due to inherent physical limitations, such as their well capacity and quantization precision~\cite{bandoh2010recent}. For example, an 8-bit sensor provides 256 intensity levels, and when a pixel exceeds its maximum voltage, signal clipping occurs, leading to a loss of detail in overexposed areas, as depicted in Figure~\ref{fig:main}(a). This imbalance between saturated and dark regions defines the camera's Dynamic Range (DR), calculated as the ratio between the maximum measurable signal (Full Well Capacity, FWC) and the lowest distinguishable signal above the noise floor (root-mean-square Dark Noise, NF). Although commercial sensors offer the ability to adjust exposure value based on DR, it is insufficient to capture the wide intensity range of scenes found in real-world environments~\cite{so2022mantissacam}.

\textcolor{black}{Existing solutions for increasing dynamic range (DR) include capturing multiple images at different exposure levels and subsequently fusing them using computational techniques~\cite{debevec2023recovering,robertson2003estimation}, or enhancing bit depth through hardware modifications~\cite{vanderperren2006approach, smaragdis2009dynamic, anderson2009analysis}. However, the DR attainable through these methods remains bounded, often leading to a trade-off between spatial resolution and DR, where crucial details of HDR content are usually lost~\cite{zhao2015unbounded,zhou2020unmodnet,so2022mantissacam}. \textcolor{black}{A promising solution arises with modulo imaging, also known as unlimited/unsaturated sensors. Modulo imaging (MI) can use an array of US-ADCs (Unlimited Sampling Analog-to-Digital Converters) that reset the signal upon reaching the saturation level,} in continuous time, before sampling~\cite{bhandari2021unlimited}. MI technology offers the potential to enhance DR capabilities by indirectly sensing an unlimited intensity range. This non-linear sensing process results in \textcolor{black}{AWGN-contaminated images with discontinuous} shapes with discontinuous shapes, as illustrated in Figure~1(b), where a reconstruction algorithm becomes necessary to obtain a smooth and continuous scene representation.}

\textcolor{black}{Several reconstruction strategies have been proposed to recover HDR signals from MI measurements, ranging across model-based, sampling-theoretic, and learning-based paradigms. Phase-unwrapping techniques adapted from optical sensing~\cite{pineda2020spud,ramirez2024phase,bacca2024deep} exploit local smoothness assumptions to infer wrap counts and recover the underlying signal. Unlimited sampling-based methods~\cite{bhandari2020unlimited,bhandari2021unlimited,guo2023iter} instead formulate MI reconstruction as the estimation of discrete residual (wrap-count) components from modulo observations, providing strong guarantees under specific noise conditions. In parallel, deep learning approaches aim to directly reconstruct HDR content from modulo measurements by learning complex image priors from data~\cite{zhou2020unmodnet,so2022mantissacam}. Despite addressing the same reconstruction task, these methods offer complementary strengths but face practical limitations. Phase-unwrapping approaches depend on smoothness priors and Itoh's condition, often violated in natural images, causing error propagation in textured or high-contrast regions. Unlimited sampling methods, though theoretically sound, are highly noise-sensitive when estimating discrete residuals, leading to strong performance drops in realistic settings. Learning-based models improve robustness but often generalize poorly to unseen noise levels, incur high computational cost, or lack interpretability. } 

\textcolor{black}{To mitigate these limitations, our previous work~\cite{bacca2024deep} introduced a Plug-and-Play ADMM fomrulation for MI reconstruction,showing that a learned denoiser used as an implicit prior can substantially outperform classical unwrapping and unlimited sampling baselines in noisy regimes. This work demonstrated the effectiveness of combining model-based optimization with data-driven priors for modulo imaging and marked an important step toward practical HDR reconstruction. However, it mainly validated the Plug-and-Play formulation and did not fully address several key practical requirements. In particular, the method relied on noise-specific denoisers with limited generalization across noise levels, incurred high runtime and memory costs due to a large number of ADMM iterations, and remained sensitive to structural assumptions that may not hold consistently across diverse natural scenes.}

\textcolor{black}{\textcolor{black}{Therefore, this work proposes an unrolled optimization framework that integrates the complementary strengths of model-based reconstruction and deep-learning priors for HDR recovery from modulo measurements corrupted by additive white Gaussian noise (AWGN). This unrolled formulation, combined with deep priors, enables the model to effectively handle local violations of Itoh's condition, improving robustness and generalization to diverse natural scenes.} Furthermore, a self-supervised Scaling Equivariance fine-tuning strategy is introduced to adapt the network to out-of-distribution modulo images. Extensive simulations demonstrate that the proposed Unrolled-MI network achieves improvements of up to 3.5 dB in PSNR compared to state-of-the-art methods across various noise scenarios, while preserving computational efficiency and scalability.}. Key contributions highlight the effectiveness and novelty of our approach:

\begin{itemize}
    \item \textcolor{black}{\textbf{Unrolled Optimization as a Trainable Network:} The classical ADMM solver is transformed into differentiable layers with learnable parameters for modulo imaging. The model learns data-driven priors that effectively capture non-Gaussian artifacts, leading to fewer iterations and enhanced reconstruction quality.}

     \item \textcolor{black}{\textbf{Lightweight Shared Network for Robust-Denoising:} A compact deep denoiser, pre-trained on various noise conditions, is included in each unrolled layer and fine-tuned jointly. It addresses diverse AWGN levels with a single model, ensuring low parameters and fast inference.}

     \item \textcolor{black}{\textbf{Self-Supervised Scaling Equivariance for Out-of-Distribution Adaptation}: The network adapts to new lighting conditions without ground-truth labels by generating virtual training pairs through random intensity   scaling and re-wrapping of HDR reconstructions. This process minimizes discrepancies between outputs, providing a self-supervised loss to reduce wrap-around artifacts.}
\end{itemize}

\section{Related Work}

\subsection{Single-Shot High-Dynamic Range Imaging}
\textit{Hardware methods} target the ADC stage to mitigate dynamic range limitations \cite{vanderperren2006approach, smaragdis2009dynamic, anderson2009analysis}. Some solutions propose a logarithmic response during photon-to-electron conversion to extend dynamic range \cite{cheng2007high}. Nevertheless, these sensors exhibit increased photon noise, especially in low-light conditions.

\textit{Computational methods} involve directly learning a reconstruction operator, which transforms a Low Dynamic Range (LDR) image map into an HDR image through a black-box inverse mapping process \cite{abel1991restoring, ting2013mitigation, esqueda2016aliasing, logan1984signals, janssen1986adaptive}. Deep models, ranging from convolutional neural networks~\cite{kalantari2017deep,eilertsen2017hdr} to transformers~\cite{liu2022ghost,shang2024hdrtransdc}, represent the state-of-the-art solutions for this problem~\cite{wang2021deep}. While these methods perform well in small and low-light scenarios, their performance degrades in high-intensity areas requiring extensive completion.

\textit{Mixed Hardware-Computational methods} combine hardware and software techniques to address the dynamic range problem. Inspired by the concept of multiple exposure times, some studies propose varying the exposure time at the pixel level \cite{nayar2000high}, employing a spatially-variant gain sensor \cite{hajisharif2015adaptive}, or designing sensors with different pixel sizes \cite{street1998high}. These approaches fuse all images at different exposure times using recovery algorithms. However, these alternatives sacrifice image resolution and are limited by a small number of possible levels. Optical-encoding elements such as cross-screen filters~\cite{rouf2011glare} or diffractive optical elements~\cite{metzler2020deep,sun2020learning} have been included in cameras to encode bright pixel values into nearby ones, enabling HDR image reconstruction through deep decoders~\cite{metzler2020deep,sun2020learning,rouf2011glare}. However, practical scenarios often result in strong spatial degradations and ill-posed reconstruction, leading to a loss of spatial resolution and image artifacts.

\begin{figure*}[!t]
    \centering
    \includegraphics[width=\linewidth]{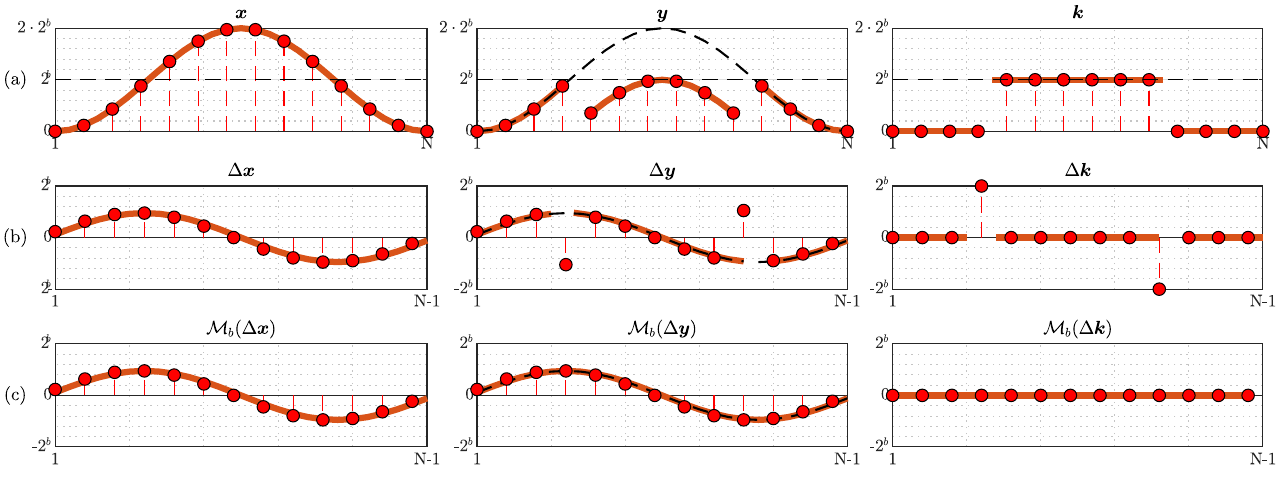}
    \caption{ (a) The signal $\boldsymbol{x}$ is decomposed as \textcolor{black}{$\boldsymbol{x = y} + \boldsymbol{k} \cdot 2^b$}. (b) \textcolor{black}{(b) The  application of spatial different operator ($\Delta$) to each term}. Notice that each term of $\Delta \boldsymbol{k}$ is 0 except for the values where the reset occurs, which have values of $\pm 2^{b}$. (c) The results of applying the centered modulo operator $\mathcal{M}_b$ to each term. Notice that $\Delta \boldsymbol{x} = \mathcal{M}_b(\Delta \boldsymbol{y})$. }
    \label{fig:wpdiff} \vspace{-1em}
\end{figure*}

\subsection{Modulo Imaging}

MI has been implemented with self-reset ADCs~\cite{zhao2015unbounded} or with programmable sensors~\cite{so2022mantissacam}. However, with the advancement of Unlimited Sampling Theory, it is now possible to implement it using recent US-ADCs in continuous time, i.e., prior sampling~\cite{bhandari2021unlimited}. Several reconstruction algorithms based on modulo imaging have been proposed for HDR images. These include adapting existing phase unwrapping algorithms for natural images~\cite{bacca2024deep}, \textcolor{black}{directly estimating the missing reset levels using the cumulative sum operator of high-order differences~\cite{bhandari2020hdr}}, using Unet-based networks for logarithmic modulo samples~\cite{so2022mantissacam}, employing iterative binary roll-over estimation using deep neural networks~\cite{zhou2020unmodnet}, \textcolor{black}{and adaptations of unlimited-sampling algorithms, such as incorporating neighborhood vectorization of pixel paths and stripe artifact removal~\cite{monroy2025autoregressive}. Nevertheless, current methods remain heavily dependent on training conditions and are sensitive to the estimation of wrapping levels, potentially resulting in suboptimal performance when exposed to high levels of noise in the modulo measurements.}

\break

\subsection{Deep Unrolling}

Deep unrolling~\cite{monga2021algorithm} systematically bridges optimization algorithms and deep neural networks, leveraging their generalization capability and interpretability~\cite{monroy2022jr2net}. By converting each iterative optimization step into a network layer, this technique creates an optimization-inspired deep neural network. In computational imaging, deep unrolling has been employed to replace the prior term with a learnable module, effectively bridging model-based optimization and deep learning-based image restoration. Unrolling models have been successfully applied to diverse inverse problems, such as compressive sensing~\cite{yang2018admm}, photon-limited image deblurring~\cite{sanghvi2022photon}, single image super-resolution~\cite{zhang2020deep}, \textcolor{black}{and recently for modulo imaging~\cite{chen2025robust}. However, a common drawback of existing unrolled models is their design for specific datasets and noise conditions, which limits adaptability to new scenes and necessitates large-scale networks with high computational overhead~\cite{chen2025robust}. Therefore, we focus on the development of lightweight unrolled models that can robustly handle variable noise levels, particularly in high dynamic range imaging from modulo measurements.}

\section{Modulo Observation Model}
\label{sec:formulation}

The sensing operation of an MI with a $b$ bit-depth in the camera recording is to reset the intensity value when it reaches a maximum threshold of $2^b$. This modulo sensing process can be mathematically modeled as
\textcolor{black}{
 \begin{equation}
\boldsymbol{y} = \mathcal{W}_b(\boldsymbol{x}) 
 = \boldsymbol{x} - 2^b \cdot \Big\lfloor \frac{\boldsymbol{x}}{2^b} \Big\rfloor
  , \label{eqn:model}
\end{equation}
}where $\boldsymbol{y}$ represents the modulo measurements, $\boldsymbol{x}$ refers to the desired HDR image, \textcolor{black}{and $\lfloor \cdot \rfloor$ represents the floor operator.} Equation~\eqref{eqn:model} represents the noiseless scenario; in this case, we can establish a relation between the HDR image $\boldsymbol{x}$ and the modulo measurements $\boldsymbol{y}$ as
\begin{equation}
\boldsymbol{x} = \boldsymbol{y} + \boldsymbol{k} \cdot 2^b,
\end{equation}
where $\boldsymbol{k}$ is an integer-value vector that represents the number of times the modulo operator has been applied for each pixel~\cite{bhandari2020unlimited} (See Figure~\ref{fig:wpdiff} (a) for a 1D visual representation of the problem). Therefore, in this scenario, the modulo inverse problem consists of detecting how many times each pixel was reset ($\boldsymbol{k}$ values) and then adding it to the measurements~\cite{bhandari2021unlimited}.  However, in a real scenario, \textcolor{black}{it is important to consider the presence of additive white Gaussian noise (AWGN)}, which affects the acquired modulo measurement. In this regard, the image formation model can be written as
\begin{equation}
\boldsymbol{y} = \mathcal{W}_b(\boldsymbol{x} + \bm{\eta}), \quad \bm{\eta} \sim \mathcal{N}(0, \sigma^2 \textbf{I} )
\label{eq:noise}
\end{equation}
\textcolor{black}{where $\bm{\eta} $ is related to the noise inherent to the image acquisition system~\cite{lang2017robust}.} The noise is reflected in the image domain, since $ \mathcal{W}_b(\boldsymbol{x} + \bm{\eta})= \mathcal{W}_b(\mathcal{W}_b(\boldsymbol{x})+ \bm{\eta})$ \textcolor{black}{is maintained (refer to supplemental material for proof).} Therefore, the presence of noise introduces additional challenges in accurately recovering the original unsaturated intensities, especially when the presence of noise causes the signal to reset.

\begin{figure*}
    \centering
    \includegraphics[width=\linewidth]{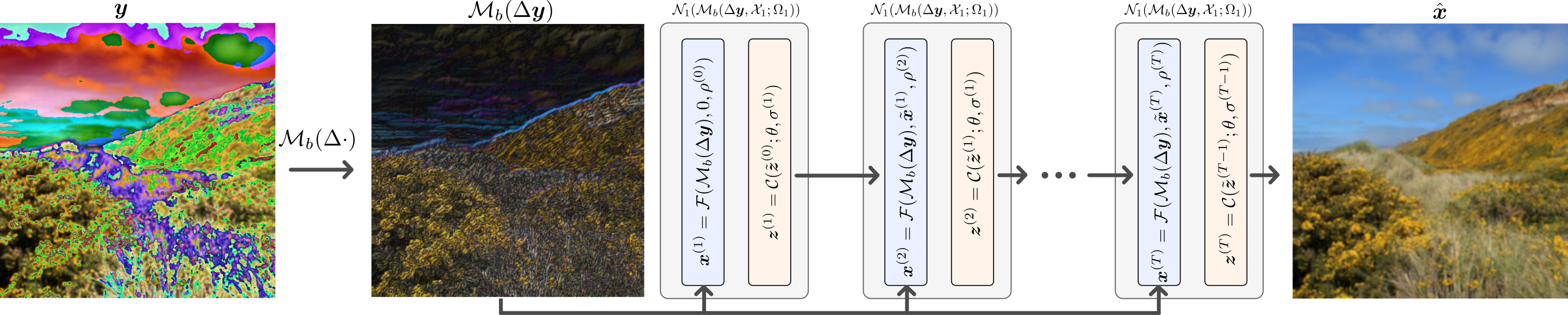}
    \caption{\textbf{Deep \textcolor{black}{Unrolling} Architecture:} The centered modulo operator is applied to the image, which is then passed through $T$ learned block layers inspired by the ADMM formulation. The HDR image is reconstructed as the output of the final layer. } \vspace{-1em}
    \label{fig:Unfolded}
\end{figure*}

\section{Deep \textcolor{black}{Unrolling} for Modulo Imaging \label{sec:method}}
\label{sec:formulation}

\textcolor{black}{Deep Unrolling merges model-driven optimization procedures with deep learning priors, enhancing efficiency and efficacy in HDR reconstruction. Specifically, the proposed approach relies on two key assumptions: firstly, the spatial finite differences in the desired HDR image, denoted as $\Delta \boldsymbol{x}$, are observed to be the same as those in the modulo measurements, except for pixels where discontinuities lead to multiples of $2^b$ (as shown in Figure 2(b))~\cite{bhandari2017unlimited}. Here, $\Delta$ represents the spatial finite difference operator, computing per-pixel differentiation in both horizontal and vertical directions. Hence, by re-wrapping the spatial differences of the measurements, we retrieve the spatial difference of the HDR image, provided $||\Delta \boldsymbol{x}||_\infty < 2^b$, as illustrated in Figure 2(c). Secondly, an implicit regularization function $\mathcal{R}(\cdot)$ is assumed, which incorporates image priors, denoising, and a custom deep prior for the problem. Based on these considerations, we can formulate the recovery problem as follows} \begin{equation}
\label{eq:main_problem}
 \boldsymbol{\hat{x}} = \underset{\boldsymbol{x}}{\mathrm{argmin}} \,  \begin{Vmatrix} \Delta\boldsymbol{x} - \mathcal{M}_b(\Delta\boldsymbol{y}) \end{Vmatrix}_{2}^{2} + \lambda  \, \mathcal{R}\!\left ( \boldsymbol{x} \right ), \end{equation}where $\mathcal{M}_b(\cdot)$ represents a centered modulo operator. Mathematically, this operator is defined as
\begin{equation}
\mathcal{M}_b(\Delta \boldsymbol{y}) = \Delta \boldsymbol{y} - 2^b \cdot \mathrm{round}\left( \Delta \boldsymbol{y} / {2^b} \right).
\end{equation}
\textcolor{black}{Here, $\text{round}(\cdot)$ denotes the rounding operator to the nearest integer. We introduce the $\mathcal{M}_b(\cdot)$ operator in the finite-difference domain because image gradients are signed and naturally symmetric around zero. This operator maps its input to the interval $[-2^{b-1}, 2^{b-1})$.}

In order to solve \eqref{eq:main_problem}, we propose to use a standard ADMM formulation, which is then unrolled into a neural network. Specifically, ADMM suggests introducing an auxiliary variable $\boldsymbol{z}$ to rewrite \eqref{eq:main_problem} into a constrained problem as:
\begin{align}
\label{eq:constrained}
 \boldsymbol{\hat{x}} &= \underset{\boldsymbol{x}}{\mathrm{argmin}} \,  \begin{Vmatrix} \Delta\boldsymbol{x} - \mathcal{M}_b(\Delta\boldsymbol{y}) \end{Vmatrix}_{2}^{2} + \lambda  \, \mathcal{R}\!\left ( \boldsymbol{z} \right ), \nonumber \\
& \text{subject to} \;\; \boldsymbol{x=z},
\end{align}
which is used to split the $\ell_2$ fidelity terms and the implicit prior to deal then with both terms independently. Specifically, the Augmented Lagrangian associated with \eqref{eq:constrained} is described as
\begin{equation}
    \mathcal{L}(\boldsymbol{x,z,u})= \begin{Vmatrix} \Delta\boldsymbol{x} - \mathcal{M}_b(\Delta\boldsymbol{y}) \end{Vmatrix}_{2}^{2} + \lambda  \, \mathcal{R}\!\left ( \boldsymbol{z} \right ) + \frac{\rho}{2}||\boldsymbol{x-z+u}||_2^2
\end{equation}
with $\boldsymbol{u}$ as the dual variable. ADMM aims to find a saddle point of $\mathcal{L}(\boldsymbol{x,z,u})$ by solving the following sequence of subproblems:
\begin{equation} \label{eq:main_x}
	\boldsymbol{x}^{\left ( k + 1 \right )} = \underset{\boldsymbol{x} \, \in \, \mathbb{R}^{n}}{\mathrm{argmin}}  \begin{Vmatrix} \Delta\boldsymbol{x} - \mathcal{M}_b(\Delta\boldsymbol{y}) \end{Vmatrix}_{2}^{2} +\frac{\rho}{2}\begin{Vmatrix} \boldsymbol{x} - \tilde{\boldsymbol{x}}^{\left ( k \right )}\end{Vmatrix}_{2}^{2}, 
\end{equation}
\begin{equation} \label{eq:mian_z}
\boldsymbol{z}^{(k+1)}=\underset{\boldsymbol\, \, {z} \, \in  \, \mathbb{R}^n}{\operatorname{argmin}}\, \, \lambda \mathcal{R}(\boldsymbol{z})+\frac{\rho}{2}\left\|\boldsymbol{z}-\tilde{\boldsymbol{z}}^{\left ( k \right )}\right\|^2,
\end{equation}

\begin{equation} \label{eq:dual_vairbale}
	\boldsymbol{u}^{(k+1)} =\boldsymbol{u}^{(k)}+\left(\boldsymbol{x}^{(k+1)}-\boldsymbol{z}^{(k+1)}\right),
\end{equation}
with $\widetilde{\boldsymbol{x}}^{\left ( k \right )} = \boldsymbol{z}^{(k)} - \boldsymbol{u}^{(k)}$ and $\widetilde{\boldsymbol{z}}^{(k)} = \boldsymbol{x}^{\left ( k + 1 \right )} +\boldsymbol{u}^{\left ( k \right )}$. The following subsection details the solution of each subproblem and how it can be unfolded into a single interpretable neuronal network to train it in an end-to-end framework.

\subsection{Spatial Finite Difference Term}

\textcolor{black}{Notably}, Equation~\eqref{eq:main_x} can be solved pixel by pixel. Abusing the notation, we can formulate an optimization problem for each spatial pixel as
 \begin{align} 
    &\boldsymbol{x}^{\left ( k + 1 \right )}_{m,n}  =  \label{eqn:eq:main_x_pixel}    \\ &\underset{\boldsymbol{x}_{m,n} }{\mathrm{argmin}}  \begin{Vmatrix} \Delta\boldsymbol{x}_{m,n} - \mathcal{M}_b(\Delta\boldsymbol{y}_{m,n}) \end{Vmatrix}_{2}^{2}   +\frac{\rho}{2}\begin{Vmatrix} \boldsymbol{x}_{m,n} - \tilde{\boldsymbol{x}}_{m,n}^{\left ( k \right )}\end{Vmatrix}_{2}^{2}.  \notag
\end{align} \textcolor{black}{Equation~\eqref{eqn:eq:main_x_pixel} has a closed-form solution by differentiating and setting it equal to zero $\frac{\partial}{\partial\boldsymbol{x}} ( \cdot) = 0$ and taking into account the properties of the linear operator $\frac{\partial}{\partial \boldsymbol{x}} (\frac{1}{2}\Vert \Delta\boldsymbol{x} \Vert_2^2) = \Delta^\top \Delta \boldsymbol{x} $, we have that} 
\begin{align} 
    \Delta^\top \left( \Delta\boldsymbol{x}_{m,n} - \mathcal{M}_b(\Delta\boldsymbol{y}_{m,n}) \right)   + \frac{\rho}{2} (\boldsymbol{x}_{m,n} - \tilde{\boldsymbol{x}}_{m,n}^{\left ( k \right)})=0.
\end{align}
Taking into account that $\Delta^\top\Delta\boldsymbol{x}_{m,n}= 4\boldsymbol{x}_{m,n} - \boldsymbol{x}_{m-1,n}  \\ \boldsymbol{x}_{m,n-1} - \boldsymbol{x}_{m+1,n} - \boldsymbol{x}_{m,n+1}$ and that a spatial shift results in a constant multiplication in the 2D Discrete Cosine Transform (DCT)~\cite{ahmed2006discrete}, denoted as $\mathcal{D}(\cdot)$, we have the following element-wise closed-form solution
\begin{equation} \label{eq:close_x}
\begin{split}
\boldsymbol{x}_{m, n}^{(k+1)} = \mathcal{D}^{-1}\left( \frac{\mathcal{D} \left( \Delta^\top \mathcal{M}_b(\Delta\boldsymbol{y})  + ( \frac{\rho}{2} )  \tilde{\boldsymbol{x}}^{(k)}\right)_{m,n}} {2(  2  + \frac{\rho}{4} -\cos(\frac{\pi m}{M})  - \cos(\frac{\pi n}{N}) ) }\right), \end{split}
\end{equation} where $M$ and $N$ represent the dimensions of the image and $\mathcal{D}^{-1}(\cdot)$ denoted the inverse 2D DCT. \textcolor{black}{This enables efficient computation of inversion for $\ell_2$-$\ell_2$ minimization in the DCT domain. A more detailed explanation of \eqref{eq:close_x} can be found in \cite{ramirez2024phase}}. Consequently, we defined this operator as a layer of the form  
\begin{equation}
    \boldsymbol{x}^{(k+1)}=\mathcal{F}(\mathcal{M}_b(\Delta\boldsymbol{y}), \tilde{\boldsymbol{x}}^{(k)} ; \rho^{(k)}),
\end{equation}
here the first two terms refer to the input variables and $\rho^{(k)}$ as learned weights.

\subsection{Deep \textcolor{black}{Learnable} Prior Sub-problem}

 The $z$-subproblem \textcolor{black}{from} Equation~\eqref{eq:mian_z} can be interpreted as a proximal operator of $\mathcal{R(\cdot)}$ over $\tilde{\boldsymbol{z}}$. However, instead of learning the function $\mathcal{R(\cdot)}$ we learn a DNN denoiser of the form
\begin{equation}
    \boldsymbol{z}^{(k+1)} = \mathcal{C}( \tilde{\boldsymbol{z}}^{(k)} ; \theta, \sigma^{(k)} ), \label{eqn:denoiser}
\end{equation} 
where $\mathcal{C}(\cdot;\cdot)$ represents a DNN with parameter $\theta$ and $\sigma^{(k)} = \sqrt{\lambda / \rho^{(k)}}$ related to a noise level inspired in PnP schemes~\cite{bacca2024deep}. In this fully differentiable operator, the input variable is $\tilde{\boldsymbol{z}}$ and the learned parameters are $\theta$ and~$\sigma^{(k)}$.

\subsection{Deep \textcolor{black}{Unrolling} Architecture}

\textcolor{black}{Deep unrolling algorithm systematically transforms iterative optimization algorithms into trainable neural network architectures by interpreting each iteration as a network layer, enabling end-to-end learning of parameters and improved convergence properties~\cite{monga2021algorithm}.} In our approach, each step of the ADMM algorithm is reformulated as a fully differentiable layer or block, with a set of learned parameters. Specifically, for the $k$-th layer, we denote the learned parameters as $\Omega_k = \{\rho^k, \theta^k, \sigma_k\}$ and the optimization variables as $\chi_k = \{\boldsymbol{u}^{(k)}, \boldsymbol{x}^{(k)}, \boldsymbol{z}^{(k)}\}$.

\textcolor{black}{Mathematically, the unrolled network can be expressed as a composition of $T$ iterative modules, each corresponding to one iteration of the original optimization algorithm:
\begin{equation}
    \hat{\boldsymbol{z}} = \mathcal{F}_T \circ \cdots \circ \mathcal{F}_1(\mathcal{M}_b(\Delta\boldsymbol{y}), \chi_1 ; \Omega_1),
    \label{eq:deep_unroll_compose}
\end{equation}}\textcolor{black}{where each $\mathcal{F}_k$ implements the update rules for the variables at iteration $k$. This composition can be compactly written as
\begin{equation}
    \hat{\boldsymbol{z}} := f(\boldsymbol{y}; \Omega),
    \label{eq:deep_prior}
\end{equation}
where $f(\cdot; \Omega)$ represents the deep unrolling network with all learnable parameters $\Omega = \{\Omega_1, \ldots, \Omega_T\}$. A visual representation of the inspired network is shown in Figure~3.} To train the network, we employ an end-to-end scheme using error back-propagation to optimize all parameters jointly. To ensure that a single model is robust to different noise levels, we propose the following optimization problem:
\begin{equation}
\label{eq:E2E}
 \hat{\Omega} \in \underset{\Omega}{\mathrm{argmin}} \hspace{0.2em}
 \mathbb{E}_{\boldsymbol{x},\bm{\eta}} \left\| \boldsymbol{x} - f( \mathcal{W}_b(\boldsymbol{x} + \bm{\eta}) ; \Omega) \right\|_{2}^2,
\end{equation}
where, in practice, the expectation is approximated using a fixed set of $\mathcal{J}$ HDR scenes, $\{\boldsymbol{x}_j\}_{j=1}^\mathcal{J}$, and different AWGN samples $\bm{\eta} \sim \mathcal{N}(0, \sigma^2 \mathbf{I})$ generated at each epoch. The noise level $\sigma$ is used as an input, serving as the hyperparameter $\lambda$ in Equation~\eqref{eq:deep_prior}. Once the network is trained, the reconstruction for a test image is obtained as $\hat{\boldsymbol{x}}_{\text{test}} = f(\boldsymbol{y}; \hat{\Omega})$.$\hat{\boldsymbol{x}}_{test}=f(\boldsymbol{y};\hat{\Omega})$.

 \subsection{Scaling Equivariance in Out of Distribution Data.} \label{ssec:se}

Unrolled models have demonstrated superior image quality and reconstruction speed compared to traditional optimization‐based restoration methods. However, their performance degrades when evaluated with measurements that significantly differ from the training distribution.

To address this limitation, we propose a self‐supervised fine‐tuning strategy for new modulo images. Our approach is inspired by the Equivariant Imaging (EI) framework introduced by Tachella et al.~\cite{chen2021equivariant, chen2023imaging} for self‐supervised linear inverse problems. In essence, the EI framework proposes that virtual measurements generated via data augmentation enable the network to learn beyond the range space of the sensing operator. According to EI framework~\cite{chen2021equivariant}, the network and the sensing operator need to be equivariant in the composition $f\circ\mathcal{W}_b$. Following this principle, we introduce the concept of Scaling Equivariance (SE) for self‐supervised modulo imaging as
\begin{equation}
    f(\mathcal{W}_b(\alpha\boldsymbol{x});\Omega)=\alpha f(\mathcal{W}_b(\boldsymbol{x});\Omega).
\end{equation}
In this formulation, an additional scaling is applied to the unrolled estimates to generate virtual modulo images and their corresponding reconstructions, thereby reducing the discrepancy between these virtual estimations. This is mathematically modeled as

\begin{equation}
    \mathcal{L}^{\alpha}_{\text{SE}}(\boldsymbol{y}, f;\Omega)=  \;\Vert \boldsymbol{x}_2 - \boldsymbol{x}_3 \Vert_2^2,
\end{equation}where the pairs $\boldsymbol{x}_2, \boldsymbol{x}_3$ are constructed as 

\begin{equation}
\begin{aligned}
        \boldsymbol{x}_2 &=  \alpha \cdot f(\boldsymbol{y};\Omega) , \quad \alpha \sim \mathcal{U}(a,b), \\
        \boldsymbol{x}_3 &= f( \mathcal{W}_b(\boldsymbol{x}_2); \Omega).
\end{aligned}
\end{equation}

In this context, $\boldsymbol{x}_2$ denotes a new HDR image with a slightly modified intensity range due to the added scaling factor. Furthermore, $\alpha$ and $\boldsymbol{x}_3$ constitute the new HDR estimations derived from the virtual modulo image sourced from $\boldsymbol{x}_2$. The motivation behind this method is based in the self-supervised learning approach for mitigating wrapping artifacts inherent in the new HDR image, as referenced in $\boldsymbol{x}_3$, which subsequently assists in mitigating the initial artifacts found in the recovery process of $\hat{\boldsymbol{x}}=f(\boldsymbol{y};\Omega)$ based on the new modulo image~$\boldsymbol{y}$. Finally, the proposed fine-tuned strategy consist in \textcolor{black}{minimizing} $\mathcal{L}^{\alpha}_{\text{SE}}$ loss on the set of new modulo images

\begin{equation}
    \Omega^* \in \underset{\Omega}{\text{arg min}} \; \mathbb{E}_{\boldsymbol{y},\alpha}  \mathcal{L}^{\alpha}_{\text{SE}}(\boldsymbol{y}, f;\Omega). \label{eqn:se}
\end{equation}


\algdef{SE}[SUBALG]{Indent}{EndIndent}{}{\algorithmicend\ }%
\algtext*{Indent}
\algtext*{EndIndent}



\section{Simulations and Results}

In this section, we provide a comprehensive evaluation of the proposed method, denoted as (\textbf{Unrolled-MI}), in the context of modulo cameras under different noise scenarios. Consequently, we compare Unrolled-MI against state-of-the-art recovery algorithms for the modulo camera. We select a deep-based unmodulo network called \textbf{UnModNet}~\cite{zhou2020unmodnet}, and our previous work, which uses denoising priors in the inversion algorithm, referred to as \textbf{PnP-UA}~\cite{bacca2024deep}. Furthermore, due to its inherent similarity to the phase unwrapping problem, we have adopted a state-of-the-art phase unwrapping algorithm to effectively handle raw modulo measurements: the simultaneous phase unwrapping and denoising (\textbf{SPUD}) algorithm~\cite{pineda2020spud} and the fast two-dimensional phase unwrapping algorithm, denoted as \textbf{Unwrapped}~\cite{herraez2002fast}. Lastly, within the modulo recovery algorithms for one-dimensional signals, we have select \textbf{AHFD}~\cite{monroy2025autoregressive} which estimate the vectorized image from high-order finite differences.


\subsection{Dataset and Implementation Details} \label{ssec:setup}
 \textbf{Dataset.} We select the UnModNet HDR collection dataset provided by the authors in~\cite{zhou2020unmodnet}. This dataset contains 592 HDR images with resolutions of $512 \times 512$ and 10 to 12 bit depth per color channel. These images cover diverse scenes, dynamic ranges, and acquisition methods, including architecture, landscapes, and portraits. \textcolor{black}{For our evaluation, all images were standardized to a fixed 10-bit depth per channel, and the corresponding modulo images were constructed at 8-bit depth, emulating the acquisition process of a modulo camera without random underexposure scaling.}

 


\begin{table*}[!t]
\renewcommand{\arraystretch}{1.2}
\centering  
\caption{\textbf{Performance Comparison for different size configuration.} We train a single Unrolled-MI network for multiple noise level, then, each column presents testing performance for a particular \textcolor{black}{AWGN} level using the same weights.} \label{tab:modelsize}
\resizebox{\textwidth}{!}{
\begin{tabular}{l|llll|rcc}
\toprule
\multirow{2}{*}{\textbf{model size}} & \multicolumn{4}{c|}{\textbf{HDR-VDP-3/PSNR-$\mu$ by AGWN level}}  
&  \multirow{2}{*}{\textbf{\#Params }}
&  \multirow{2}{*}{\textbf{ Memory[Mb] }}
&  \multirow{2}{*}{\textbf{ Time[ms] }}   \\ \cline{2-5} 
                                     & \textbf{0}  & \textbf{20} & \textbf{40}  & \textbf{80} &  &\\ \midrule
Unrolled-MI[small]       &  8.11/20.54 &  7.93/16.47 & 7.67/14.76 &  6.96/11.88  & 127.74K & 78.57 &  44.18 \\
Unrolled-MI[medium]     &   8.61/22.42 &  8.42/19.10 & 8.13/17.37  & 7.26/14.26 & 510.46K & 106.67 &  49.87 \\
Unrolled-MI[large]     &    8.59/22.97 &  8.37/19.93 & 8.08/18.48  & 7.27/15.57 & 2040.82K & 160.15 & 53.49 \\ \bottomrule
\end{tabular}
} \vspace{-1em}
\end{table*}

\newpage

\textbf{Metrics.} \textcolor{black}{To assess the quality of the unwrapped restoration, we employed a set of six selected metrics: Q-index, peak signal-to-noise ratio (PSNR-L), and structural similarity index (SSIM-L) in the linear domain, while PSNR-$\mu$ and MS-SSIM-$\mu$, in the encoding domaing PU21~\cite{azimi2021pu21} and finally, the HDR-VDP-3 quality and constrasts distortion metric~\cite{mantiuk2023hdr}.} The Q-index~\cite{wang2002image} is a metric commonly utilized in phase unwrapping algorithms to determine the accuracy of the unwrapped phase. On the other hand, PSNR and SSIM are widely recognized image quality metrics that measure the fidelity and \textcolor{black}{structural} similarity of the restored image, \textcolor{black}{we further assess the performance using the PU21 encoding to ensure an equitable comparison of HDR quality. Additionally, HDR-VDP-3 is employed to simulate human perception quality}. The mentioned metrics were implemented in accordance with the guidelines provided in ~\cite{pineda2020spud, azimi2021pu21, mantiuk2023hdr}. To enhance the visualization of the image, a Reinhard tone mapping~\cite{reinhard2023photographic} is employed, ensuring that the HDR content is effectively presented on standard display devices.


\textbf{Training Setup.} Given the computational complexity of the Unrolled-MI network training and the challenges posed by varying noise levels, we adopt a two-phase training strategy. Initially, we pretrain the denoising module as an independent network to establish robust denoising priors across a wide range of degradation levels. This initialization step provides a strong foundation that facilitates a subsequent end-to-end fine-tuning of the full Unrolled-MI network.

\textit{Phase 1: Robust Denoising Initialization.} In this phase, we train the DRUnet denoiser (referenced in Equation~\eqref{eqn:denoiser}) independently. The network is trained with handling additive Gaussian noise, with standard deviation values uniformly sampled from the interval [0, 80]. Training is performed using an L1 loss function to minimize the difference between clean and denoised images. We employ the Adam optimizer with an initial learning rate of $10^{-3}$, a batch size of 32, and 2000 training epochs. This phase enables the denoiser to learn robust priors that are crucial for effective noise suppression in the later stages.

\textit{Phase 2: Unrolled-MI Network Fine-Tuning.}
In the second phase, the pretrained denoiser is integrated into the complete deep unrolled network $f(\cdot, \Omega)$, unrolled over $T=3$ iterations. All components of the network, including the denoising module and the data-fidelity sub-problems, are fine-tuned jointly in an end-to-end manner. To ensure stable convergence during this phase, the learning rate is reduced to $5\times10^{-5}$, and training continues for 10,000 more epochs with a batch size of 16. \textcolor{black}{AWGN} levels are again sampled from the range [0, 80] to maintain robustness throughout the fine-tuning process.

The total training duration for both phases is approximately 8 hours on a single NVIDIA RTX 3090 GPU paired with an AMD R5 5700X CPU. A complete PyTorch implementation, along with detailed configurations for both training stages, is publicly available on GitHub.\\ \texttt{\url{https://github.com/bemc22/ModuloUnrolled}}

\begin{table*}[!h]  \small
 \centering
\caption{\textbf{Quantitative performance on the UnModNet dataset.} Detailed comparison of \textcolor{black}{HDR quality assessment metrics for Unrolled-MI in [medium] variant} and competing methods across multiple \textcolor{black}{AWGN} noise levels ($\sigma \in \{0, 25, 40, 80\}$). \label{tab:quant_comp2}} 
\resizebox{\textwidth}{!}{ 
\begin{centering}
    \begin{tabular}{l|c|cccccc} 
    \toprule 
    $\sigma$ & Metrics  & Unrolled-MI[\textcolor{black}{medium}]    & Unmodnet~\cite{zhou2020unmodnet} & AHFD~\cite{monroy2025autoregressive} & PnP-UA~\cite{bacca2024deep}   & SPUD~\cite{pineda2020spud}  & Unwrapped~\cite{herraez2002fast}  \\ \midrule \midrule
     & HDR-VDP-3          &  \textbf{8.61$\pm$1.15} & 7.52$\pm$1.52 & 4.60$\pm$0.86  & 4.86$\pm$0.87 & 4.97$\pm$0.86  &  5.45$\pm$1.65 \\
     & PSNR-$\mu$     &  \textbf{22.42$\pm$3.99} & 20.70$\pm$5.83 & 2.48$\pm$4.21  & 3.61$\pm$4.73 & 3.00$\pm$4.65  & 1.89$\pm$4.10 \\
   0 & MS-SSIM-$\mu$  &  0.910$\pm$0.07 & \textbf{0.968$\pm$0.06} & 0.532$\pm$0.14  & 0.579$\pm$0.17 & 0.539$\pm$0.16 & 0.389$\pm$0.15 \\
     & PSNR-L     & \textbf{32.41$\pm$8.86} & \underline{28.92$\pm$7.47} & 27.45$\pm$10.6 & 26.76$\pm$10.3  & 26.75$\pm$10.3 & 24.62$\pm$8.81  \\
     & SSIM-L     & \textbf{0.865$\pm$0.11} & \underline{0.778$\pm$0.18} & 0.597$\pm$0.28 & 0.563$\pm$0.29  & 0.563$\pm$0.29 & 0.540$\pm$0.37  \\
     & Q-index  & \textbf{0.669$\pm$0.16} & \underline{0.555$\pm$0.18} & 0.320$\pm$0.25 & 0.287$\pm$0.24  & 0.295$\pm$0.25 & 0.213$\pm$0.21  \\ \midrule
     & HDR-VDP-3          &  \textbf{8.42$\pm$1.05} & 7.31$\pm$1.66 & 4.48$\pm$0.68  & 4.88$\pm$1.01 & 4.84$\pm$0.75 & 5.20$\pm$1.42 \\
     & PSNR-$\mu$     & \textbf{19.10$\pm$4.47} &  12.81$\pm$6.92 & 3.29$\pm$4.07 & 3.75$\pm$4.62  & 2.80$\pm$4.60 & 1.91$\pm$4.28       \\
  25 & MS-SSIM-$\mu$  & 0.723$\pm$0.17 &  \textbf{0.737$\pm$0.18} & 0.564$\pm$0.13 & 0.564$\pm$0.16  & 0.518$\pm$0.15 & 0.385$\pm$0.15    \\
     & PSNR-L     & \textbf{32.34$\pm$8.69} & \underline{28.06$\pm$6.82} & 26.20$\pm$8.92 & 26.65$\pm$10.1  & 25.84$\pm$8.74 & 23.82$\pm$7.69  \\
     & SSIM-L     & \textbf{0.864$\pm$0.10} & \underline{0.717$\pm$0.19} & 0.560$\pm$0.26 & 0.559$\pm$0.29  & 0.536$\pm$0.27 & 0.492$\pm$0.33  \\
     & Q-index  & \textbf{0.669$\pm$0.17} & \underline{0.496$\pm$0.22} & 0.268$\pm$0.20 & 0.277$\pm$0.23  & 0.250$\pm$0.20 & 0.170$\pm$0.18   \\ \midrule
     & HDR-VDP-3          &  \textbf{8.13$\pm$1.07} & 6.92$\pm$1.48 & 4.40$\pm$0.65  & 4.84$\pm$1.03 & 4.67$\pm$0.73 & 5.10$\pm$1.34 \\
     & PSNR-$\mu$     &  \textbf{17.37$\pm$4.71} & 10.94$\pm$6.69 & 3.19$\pm$4.20  & 3.82$\pm$4.50 & 2.76$\pm$4.61 &  1.91$\pm$4.28   \\
  40 & MS-SSIM-$\mu$  &  0.662$\pm$0.18 & \textbf{0.682$\pm$0.19} & 0.542$\pm$0.13  & 0.555$\pm$0.16 & 0.507$\pm$0.15 & 0.384$\pm$0.14   \\
     & PSNR-L     & \textbf{32.19$\pm$8.42} & \underline{26.92$\pm$5.97} & 25.53$\pm$7.92 & 26.56$\pm$9.93  & 25.20$\pm$7.83 & 23.52$\pm$7.06  \\
     & SSIM-L     & \textbf{0.859$\pm$0.10} & \underline{0.643$\pm$0.19} & 0.519$\pm$0.24 & 0.553$\pm$0.29  & 0.494$\pm$0.25 & 0.475$\pm$0.29  \\
     & Q-index  & \textbf{0.647$\pm$0.17} & \underline{0.439$\pm$0.21} & 0.223$\pm$0.16 & 0.263$\pm$0.22  & 0.215$\pm$0.17 & 0.163$\pm$0.16   \\ \midrule
     & HDR-VDP-3          &  \textbf{7.26$\pm$1.20} & 5.33$\pm$1.269 & 3.82$\pm$0.69  & 4.68$\pm$0.98 & 4.18$\pm$0.76 & 4.63$\pm$1.09  \\
     & PSNR-$\mu$     &  \textbf{14.26$\pm$4.40} & 7.40$\pm$6.14 &  2.65$\pm$4.17  & 3.83$\pm$4.44 & 2.54$\pm$4.52 &  1.75$\pm$4.06  \\
  80 & MS-SSIM-$\mu$  &  0.516$\pm$0.19 & \textbf{0.573$\pm$0.19} & 0.465$\pm$0.12  & 0.528$\pm$0.15 & 0.468$\pm$0.14 & 0.366$\pm$0.14   \\
     & PSNR-L     & \textbf{30.99$\pm$7.91} & 22.65$\pm$3.99 & 23.57$\pm$6.39 & \underline{26.21$\pm$9.44}  & 23.72$\pm$6.18 & 22.23$\pm$5.96  \\
     & SSIM-L     & \textbf{0.826$\pm$0.12} & 0.360$\pm$0.17 & 0.356$\pm$0.17 & \underline{0.528$\pm$0.28}  & 0.354$\pm$0.18 & 0.336$\pm$0.25  \\
     & Q-index  & \textbf{0.573$\pm$0.19} & 0.250$\pm$0.17 & 0.122$\pm$0.09 & \underline{0.208$\pm$0.18}  & 0.133$\pm$0.11 & 0.097$\pm$0.11   \\ \midrule
    \multicolumn{2}{c|}{Time [ms]} 
    & 49.87 & 5327.32 &   8.34 & 456.59 &  5.51 & 222.01
    \\ 
    \multicolumn{2}{c|}{GFLOPs} 
    & 14.03 & 1245.23 &  0.05 & 108.73 & 0.03 & -
    \\
    \bottomrule
    \end{tabular}
\end{centering}
}  \vspace{-1em}
\end{table*}

\break

\subsection{Model Size Evaluation.}  \textcolor{black}{We evaluated the proposed architecture in different model size configurations to balance performance, computational efficiency, and feasibility of deployment. The baseline DRUnet architecture employs a hierarchical five-level design with convolutional filters [64, 128, 256, 512] per level, resulting in approximately 4.2M trainable parameters. To optimize resource utilization, we systematically reduced the number of filters at each level, yielding three lightweight variants: small (initial filters: 8), medium (16), and large (32). These configurations decrease the number of parameters by 93.7\%, 75.0\%, and 51.4\%, respectively, compared to the baseline.}

\textcolor{black}{As shown in Table~\ref{tab:modelsize}, the small variant achieves efficient performance with 127.74K parameters and 78.57\,MB of VRAM, allowing real-time inference at 44.18\,ms. Its performance, evaluated using HDR-VDP-3 and PSNR-$\mu$, gradually declines with increasing noise: 8.11/20.54 ($\sigma=0$), 7.93/16.47 ($\sigma=20$), 7.67/14.76 ($\sigma=40$), and 6.96/11.88 ($\sigma=80$). This indicates that, while the small model is lightweight and fast, its ability to handle high noise levels is more limited compared to the larger variants.}

The medium and large models show improved robustness to noise. The medium variant maintains higher scores across all AWGN levels, reaching 8.61/22.42 ($\sigma=0$), 8.42/19.10 ($\sigma=20$), 8.13/17.37 ($\sigma=40$), and 7.26/14.26 ($\sigma=80$). The large variant achieves comparable results of 8.59/22.97 ($\sigma=0$), 8.37/19.93 ($\sigma=20$), 8.08/18.48 ($\sigma=40$), and 7.27/15.57 ($\sigma=80$). \textcolor{black}{The parameter count of these architectures increases quadratically with the base number of feature channels. Thus, lowering the base number of feature channels from the baseline significantly reduces both parameters and memory usage.}

\textcolor{black}{Given its strong performance, within 0.5 HDR-VDP-3 units and a few PSNR-$\mu$ points of the large model across all AWGN levels, combined with substantially lower memory and parameter requirements, we selected the medium variant for all subsequent experiments.}

\begin{figure*}[!t]
    \centering
    \includegraphics[width=\linewidth]{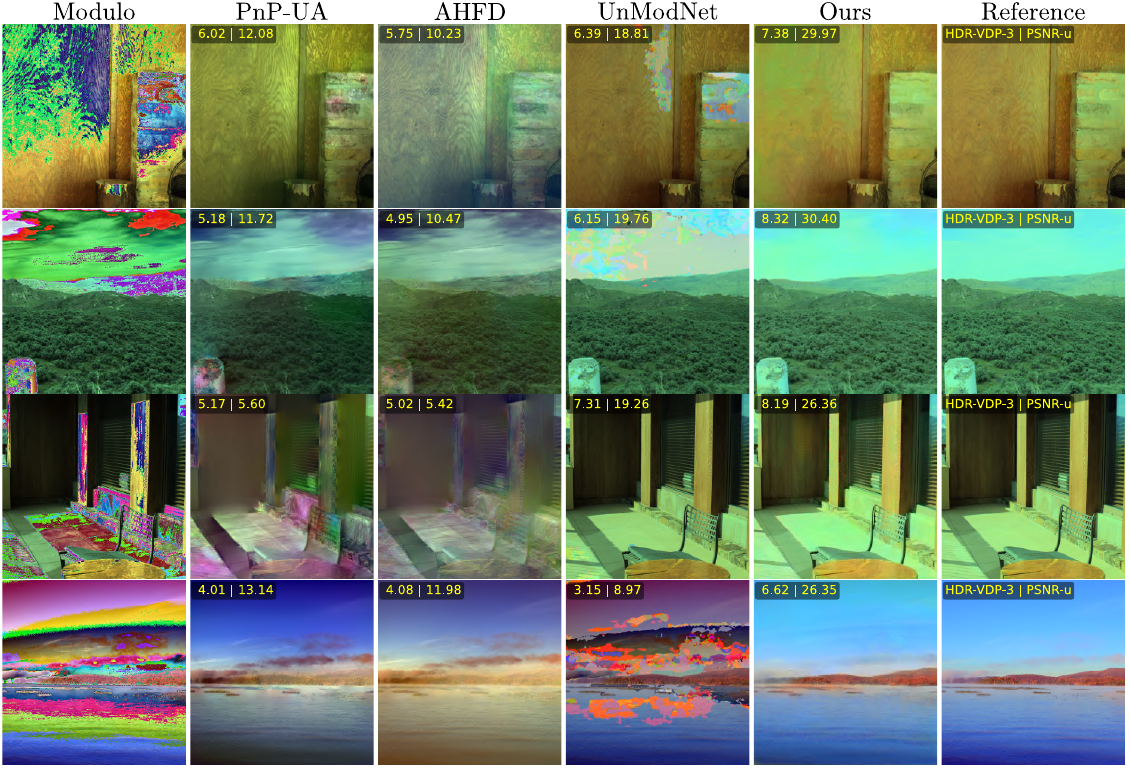} \vspace{-1.5em}
    \caption{\textbf{Visual Results of HDR Recovery Methods.} Qualitative comparison of unwrapped images from various methods, demonstrating the proposed Unrolled approach achieves superior visual fidelity and quantitative metrics (PSNR$\vert$SSIM) compared to state-of-the-art alternatives. Visual HDR images are displayed using Reinhard tone mapping on luminance component~\cite{reinhard2023photographic}.}
    \label{fig:visual_results} \vspace{-1em}
\end{figure*}
\subsection{State-of-the-art Comparison.}

\textcolor{black}{We evaluated the performance of each method under various AWGN scenarios with standard deviations of $\sigma \in \{ 0, 25, 40, 80\}$. The results, reported in Table~\ref{tab:quant_comp2}, demonstrate that our proposed Unrolled-MI method outperforms competing approaches across all noise levels when assessed in the HDR-VDP-3, PSNR-$\mu$, PSNR-L, SSIM-L and Q-index metrics.}

\textcolor{black}{Specifically, at zero noise, Unrolled-MI achieves an HDR-VDP-3 score of $8.61 \pm 1.15$, surpassing UnModNet ($7.52 \pm 1.52$) and all other methods by a considerable margin. Under light noise ($\sigma=25$), Unrolled-MI maintains a leading HDR-VDP-3 of $8.42 \pm 1.05$, compared to $7.31 \pm 1.66$ for UnModNet. Even at moderate noise ($\sigma=40$), Unrolled-MI achieves $8.13 \pm 1.07$, consistently outperforming UnModNet ($6.92 \pm 1.48$), and retains stable performance even at high noise levels ($\sigma=80$) with $7.26 \pm 1.20$, whereas UnModNet declines to $5.33 \pm 1.27$. \textcolor{black}{We recommend that the reader consult the supplementary materials for an in-depth analysis, where a histogram evaluation of the metrics score is presented.}}

\newpage

\textcolor{black}{However, SPUD, a non-iterative formulation of Unrolled-MI employing the hard-thresholding proximal operator, exhibits a considerable performance drop, particularly under higher AWGN. At ($\sigma=80$), the HDR-VDP-3 score for SPUD decreases to $4.18 \pm 0.76$, which is substantially lower than the $7.26 \pm 1.20$ achieved by Unrolled-MI. This degraded perceptual performance stems from the limitations of hard thresholding, which is less effective at preserving fine image details and mitigating strong noise artifacts. In contrast, UnModNet, as an iterative approach, incurs increased computational costs due to its reliance on binary mask estimation for wraps. Unrolled-MI, with its unrolling architecture, effectively reduces this overhead while consistently providing superior reconstruction quality at every noise level. This design allows Unrolled-MI to robustly unwrap and denoise the input image within each stage, resulting in notably higher HDR-VDP-3 scores and improved visual fidelity compared to both SPUD and UnModNet, especially in challenging high-noise scenarios.}

\textcolor{black}{In Figure~5, we show visually and quantitatively the performance of Unrolled-MI for HDR recovery. Across diverse scenes, Unrolled-MI consistently produces reconstructions visually closest to the ground truth, with minimal artifacts, well-preserved detail, and accurate tonal reproduction. This visual superiority directly correlates with the observed gains in HDR-VDP-3 and PSNR-$\mu$. This strong stability and higher overall HDR-VDP-3 scores across all noise conditions highlight the robustness and effectiveness of the proposed Unrolled-MI architecture for HDR reconstruction from modulo images. Compared to competing methods, Unrolled-MI delivers consistently higher perceptual quality, demonstrating resilience against noise-induced degradations. }

\subsection{Out of distribution evaluation.}

Our Unrolled-MI was evaluated on out-of-distribution images. Specifically, we utilized the Real RGB dataset from Unmdonet paper~\cite{zhou2020unmodnet}, which comprises ten real RGB images captured with a Fujifilm X-T20 mirrorless digital camera. These images were intensity-rescaled to a 10-bit depth per channel and captured as modulo images with a threshold of 255. To tailor the Unrolled-MI network to these novel images, we implemented the self-supervised fine-tuning approach detailed in Subsection~\ref{ssec:se} by minimizing the Scaling Equivariance loss as described in Equation~\eqref{eqn:se}. This process was conducted over 1000 epochs with an initial learning rate of $10^{-5}$, with data augmentation consisting of random horizontal and vertical flips applied to the modulo images.

\begin{table}[!h] \centering
\caption{Out-of-distribution evaluation results.} 
\label{table:3}
\vspace{-0.5em}
\resizebox{0.5\linewidth}{!}{
\begin{tabular}{l|ccc}
\toprule
Method              & PSNR-L & SSIM-L &  Q-index \\ \midrule  \midrule
Unwrapped~\cite{herraez2002fast}        & 22.24 & 0.495 & 0.307 \\
SPUD~\cite{pineda2020spud}              & 26.64 & 0.542 & 0.421 \\
PnP-UA~\cite{bacca2024deep}             & 26.70 & 0.541 & 0.421 \\
AHFD~\cite{monroy2025autoregressive}  & 27.89 & 0.646 & 0.504 \\
UnModNet~\cite{zhou2020unmodnet}        & 28.04 & 0.741 & 0.618 \\ 
Unrolled-MI           & \underline{29.94} & \underline{0.816} & \underline{0.667} \\
Unrolled-MI + SE   &     \textbf{30.75} & \textbf{0.803} & \textbf{0.667} \\ \bottomrule
\end{tabular}
} \vspace{-2em}
\end{table}

\break 

\begin{figure*}[!t]
    \centering
    \includegraphics[width=\linewidth]{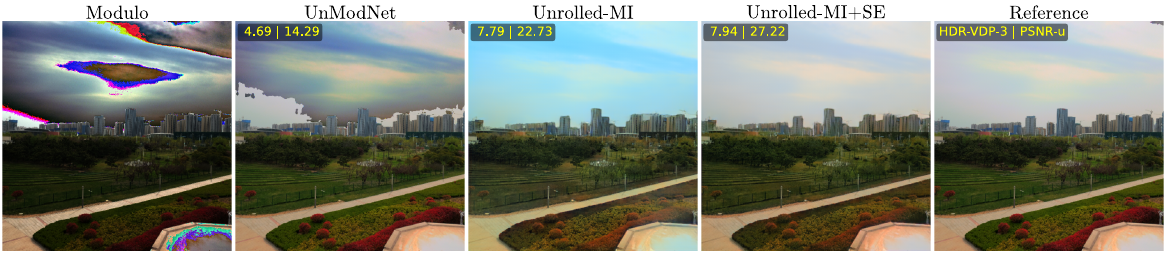} \vspace{-1.5em}
    \caption{\textbf{Out of distribution evaluation.} Visual comparison of Unrolled and UnModNet on real RGB captures}
    \label{fig:outdistri} \vspace{-1em}
\end{figure*}

Visual results are presented in Figure~\ref{fig:outdistri}. The Modulo column displays the wrapped input, while the UnModNet and Unrolled-MI columns show the unwrapped results, with the \textcolor{black}{reference} image provided for comparison. Qualitatively, UnModNet exhibits noticeable artifacts and color distortions across different scenes, whereas Unrolled-MI consistently recovers images that closely resemble the ground truth, demonstrating significantly improved artifact suppression and detail preservation. Table~\ref{table:3} provides a quantitative analysis. Results indicate that the proposed Unrolled-MI, particularly when enhanced with SE fine-tuning, achieves superior PSNR-L, SSIM-L, and Q-index values compared to baseline methods.

\begin{figure*}[h]

\newcommand{\insetFig}[1]{ \includegraphics[width=0.3\linewidth]{#1}  }
\setlength{\tabcolsep}{0pt}

    \centering

\newcommand{\imgWidth}{0.325\linewidth}
\newcommand{\h}{0.08\textwidth}
\newcommand{\w}{\imgWidth - .001\textwidth}

\newcommand{\zx}{-0}
\newcommand{\zy}{0.9}

\newcommand{\zxx}{-1}
\newcommand{\zyy}{-1}

\newcommand{\off}[1]{6*#1}

\newcommand{\bb}{2.1}
\newcommand{\ba}{-2.7}
\newcommand{\bc}{0}

 \newcommand{\addimage}[3]{
\node at (\off{#1}, 0)(img){\includegraphics[width=\imgWidth]{realdata/#2}};
\spy [green, height=\h, width=\w] on (\zx+\off{#1},\zy) in node at (\off{#1},\ba);
\node at (\off{#1}, \bb) {#3};
}

 \newcommand{\addimagee}[2]{
\node at (\off{#1}, 0)(img){\includegraphics[width=\imgWidth]{realdata/#2}};
\spy [green, height=\h, width=\w] on (\zxx+\off{#1},\zyy) in node at (\off{#1},\ba);
}

    \begin{tikzpicture}[outer sep=0pt,inner sep=1pt]\tikzstyle{every node}=[font=\small]
    
    \begin{scope}[node distance = 0mm, inner sep = 0pt, outer sep = 1,spy using outlines={rectangle, green, magnification=4.5, every spy on node/.append style={thick}}]

    \addimage{0}{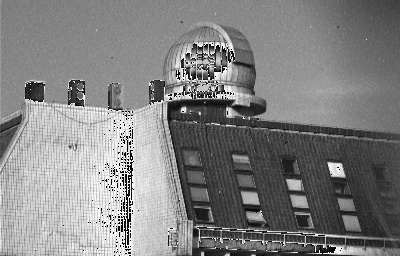}{Modulo Image}
    \addimage{1}{img1_UnModNet}{UnModNet~\cite{zhou2020unmodnet}}
    \addimage{2}{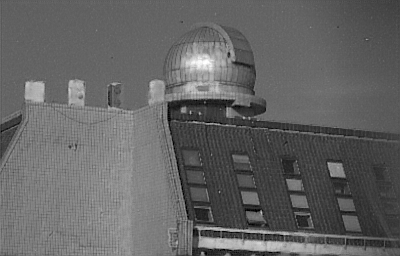}{Unrolled-MI+SE}

    \end{scope}

    \end{tikzpicture}
    
    \caption{\textbf{Real modulo data evaluation.} Comparison of HDR reconstructions from real modulo sensor captures, contrasting the performance of UnModNet and the proposed Unrolled-MI+SE method.}
    \label{fig:realdata}
    
\end{figure*}
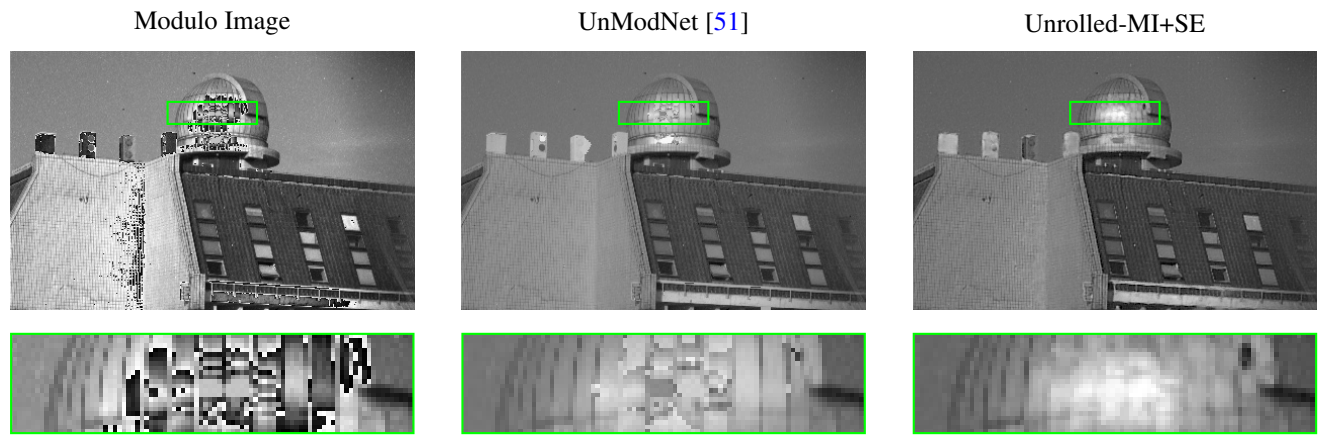

\subsection{Real modulo data evaluation.}

The Unrolled-MI was evaluated using the real modulo data set provided by the authors in~\cite{zhou2020unmodnet}. The fine-tuning procedure was followed to minimize the loss of scaling equivariance, as detailed in Equation~\eqref{eqn:se}. This process was implemented over 100 epochs, starting with an initial learning rate of $10^{-5}$. Data augmentation was achieved through the application of random horizontal and vertical flips on modulo images. Given that the real modulo data comprise a single channel, the RGB version of the Unrolled-MI network was fine-\textcolor{black}{tuned} on a grayscale version of the UnModNet dataset, replicating the grayscale channel to reuse the RGB weights, employing the methodology outlined in Section~\ref{ssec:setup}.

In Figure~\ref{fig:realdata} we present HDR image reconstruction from UnModNet and Unrolled-MI+SE. Unrolled-MI+SE demonstrates significantly improved recovery by effectively eliminating wrap-around artifacts and restoring continuous intensity gradients in saturated regions. In contrast, UnModNet output still exhibits residual discontinuities and localized distortions, especially along sharp edges. However, very high-noise areas, such as specular highlights and deep shadows, remain challenging for both approaches, with occasional erroneous resets. Thanks to its lightweight, unrolled architecture and self-supervised Scaling Equivariance fine-tuning, Unrolled-MI achieves this enhanced fidelity while maintaining competitive inference times and significantly fewer parameters than previous deep iterative algorithms.

\section{Conclusions}

In this work, we presented a deep unrolled network for high-dynamic range modulo imaging that recovers continuous intensity values from modulo measurements under varying noise conditions. By unrolling an ADMM-based optimization algorithm that exploits spatial differences via a closed-form Discrete Cosine Transform formulation, our method achieves rapid reconstruction with a lightweight architecture. The integration of a learned denoiser within the algorithm unrolling scheme effectively mitigates noise, while the integration of Scaling Equivariance term enables self-supervised fine-tuning to adapt the model to new, out-of-distribution modulo images. Extensive simulations demonstrate that our approach consistently improves performance (in some cases by up to 3.5 dBs in PSNR relative to state-of-the-art techniques), while maintaining computational efficiency and scalability. These results highlight the potential of unrolling models as a robust and adaptable solution for HDR imaging in practical applications.


\section*{Acknowledgment}

The authors extend their acknowledgment to Chu Zhou et al.~\cite{zhou2020unmodnet} for providing support by sharing the dataset of the UnModNet model and Valentina Hincapie-Castillo for providing beautiful HDR images for demonstration purposes.

{\small
\bibliographystyle{ieee_fullname}
\bibliography{egbib}
}

\newpage

\begin{center}
    \section*{Deep Lightweight Unrolled Network for \\
High Dynamic Range Modulo Imaging\\(Supplemental Materials)}
\end{center}

\noindent This supplementary material is divided into three parts: (i) proof of noise modeling, (ii) additional experimental details and results, and iii) discussion and future work.

\setcounter{section}{0}
\section{Proof for AWGN noise in Modulo Imaging.}

For any pair $x, y \in \mathbb{R}$ and a positive integer $b$, our goal is to show \begin{equation} \label{eqn:defmoudlo}
    \mathcal{W}_b(x+y) = \mathcal{W}_b\big( \mathcal{W}_b(x) + y \big)
\end{equation}
Remark the definition of $\mathcal{W}_b(x) = x - 2^b \Big\lfloor \frac{x}{2^b} \Big\rfloor\ $. Starting from the right part of Equation~\eqref{eqn:defmoudlo} and using the definition of modulo of $\mathcal{W}_b(\cdot)$, we have
\begin{equation}
    \mathcal{W}_b(\mathcal{W}_b(x) + y) =
    \mathcal{W}_b(x) + y - 2^b \Big\lfloor \frac{\mathcal{W}_b(x) + y}{2^b} \Big\rfloor\,.
\end{equation} Now, by adding and subtracting $x$ on the right-hand side:
\begin{equation}
\begin{aligned}
    \mathcal{W}_b(\mathcal{W}_b(x) + y)
    &= \mathcal{W}_b(x) + y - 2^b \Big\lfloor \frac{\mathcal{W}_b(x) + y}{2^b} \Big\rfloor + x - x \\
    &= x + y + (\mathcal{W}_b(x) - x) - 2^b \Big\lfloor \frac{\mathcal{W}_b(x) + y}{2^b} \Big\rfloor\,.
\end{aligned}
\end{equation} From the above, note that $\mathcal{W}_b(x) - x = -2^b \Big\lfloor \frac{x}{2^b} \Big\rfloor$. Substituting gives:
\begin{equation} \label{eqn:eq1}
    \mathcal{W}_b(\mathcal{W}_b(x) + y)
    = x + y - 2^b \left( \Big\lfloor \frac{x}{2^b} \Big\rfloor + \Big\lfloor \frac{\mathcal{W}_b(x) + y}{2^b} \Big\rfloor \right)\,.
\end{equation} Next, the analysis focuses on the term $\mathcal{W}_b(x)+y$ in Equation~\eqref{eqn:eq1}. According to the definition,
\begin{equation}
    \mathcal{W}_b(x) + y = x - 2^b \Big\lfloor \frac{x}{2^b} \Big\rfloor + y\,,
\end{equation}
and thus,
\[
\Big\lfloor \frac{\mathcal{W}_b(x) + y}{2^b} \Big\rfloor = \Big\lfloor \frac{x + y}{2^b} - \Big\lfloor \frac{x}{2^b} \Big\rfloor \Big\rfloor\,.
\] Since $\Big\lfloor \frac{x}{2^b} \Big\rfloor \in \mathbb{Z}$, for any $t \in \mathbb{R}$ and $n \in \mathbb{Z}$, it holds that $\lfloor t- n \rfloor = \lfloor t \rfloor - n$. Therefore,
\begin{equation} \label{eqn:eq2}
\Big\lfloor \frac{\mathcal{W}_b(x) + y}{2^b} \Big\rfloor = 
    \Big\lfloor \frac{x + y}{2^b} - \Big\lfloor \frac{x}{2^b} \Big\rfloor \Big\rfloor = \Big\lfloor \frac{x+y}{2^b} \Big\rfloor - \Big\lfloor \frac{x}{2^b} \Big\rfloor\,.
\end{equation} Substituting \eqref{eqn:eq2} in \eqref{eqn:eq1}, we have:
\begin{equation}
\begin{aligned}
    \mathcal{W}_b(\mathcal{W}_b(x) + y)
    &= x + y - 2^b \left(  \Big\lfloor \frac{x}{2^b} \Big\rfloor + \Big\lfloor \frac{x+y}{2^b} \Big\rfloor - \Big\lfloor \frac{x}{2^b} \Big\rfloor  \right) \\
    &= x + y - 2^b \Big\lfloor \frac{x+y}{2^b} \Big\rfloor \\
    &= \mathcal{W}_b(x + y)\,. \qquad \square
\end{aligned}
\end{equation}

\newpage

\section{Additional Experimental Details and Results}

\textbf{Reinhard Tone Mapping.} We present the Reinhard tone mapping method utilized for visualizing HDR estimation images. In our experiments, we employ the Reinhard global operator applied to the luminance domain~\cite{reinhard2023photographic}. For each HDR image $\mathbf{x}$, we first apply an exposure adjustment with a scalar parameter $\alpha$, so that $\mathbf{x}' = \alpha \cdot \mathbf{x}$. The luminance channel $L$ is computed using ITU-R BT.709 coefficients:
\begin{equation}
L = 0.2126R + 0.7152G + 0.0722B
\end{equation}

The Reinhard tone-mapping curve is then applied to the luminance: $L_{\text{tm}} = \frac{L}{L + \beta}$ where $\beta$ is a positive parameter controlling the dynamic range compression. To maintain color consistency, a per-pixel scaling factor is computed as $s = \frac{L_{\text{tm}}}{L + \epsilon}$ with $\epsilon$ a small constant to avoid division by zero. The final tone-mapped image is $\mathbf{x}_{\text{tm}} = \mathbf{x}' \cdot s$.

In our visualizations, different values of $\beta$ were set for each image to achieve the best visual appearance and clarity. The underlying tone-mapping function remains the same, but the choice of $\beta$ is adapted per image to enhance visual quality.

\begin{figure*}[!h]
    \centering
    \includegraphics[width=\linewidth]{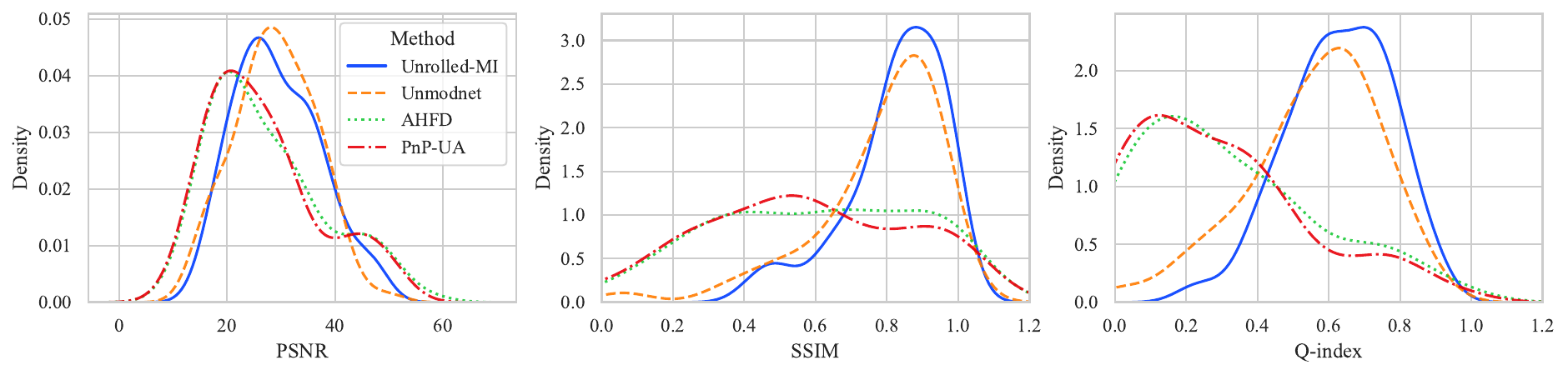} \vspace{-2em}
    \caption{\textbf{Histogram of performance metrics.} Distributions of PSNR, SSIM, and Q-index scores across various methods highlight the consistent and superior performance of the proposed Unrolled-MI approach.} 
    \label{fig:kde-scores}
\end{figure*}

\textbf{In distribution evaliation.} Figure~\ref{fig:kde-scores}, the visualizations of the histogram show that Unrolled-MI consistently outperforms alternative methods across all metrics evaluated (PSNR, SSIM, and Q-index). The right-shifted and compactly distributed density curves for Unrolled-MI demonstrate not only superior average performance but also greater stability and reliability relative to state-of-the-art methods including UnModNet, PnP-UA, SPUD, Unwrapped, and AHFD. This visual analysis strongly supports the quantitative results, highlighting the robustness and effectiveness of the proposed Unrolled-MI architecture for HDR reconstruction from modulo images.

\textbf{Out of distribution evaluation.} To assess the robustness and generalization capabilities of the proposed HDR reconstruction methods, we perform an out-of-distribution evaluation using the 10 real RGB scenes from the UnModNet dataset. This evaluation is designed to rigorously test each method on challenging, previously unseen content captured with modulo cameras, representing diverse indoor and outdoor conditions.

\begin{figure}[!h]
        \centering
        \includegraphics[width=\linewidth]{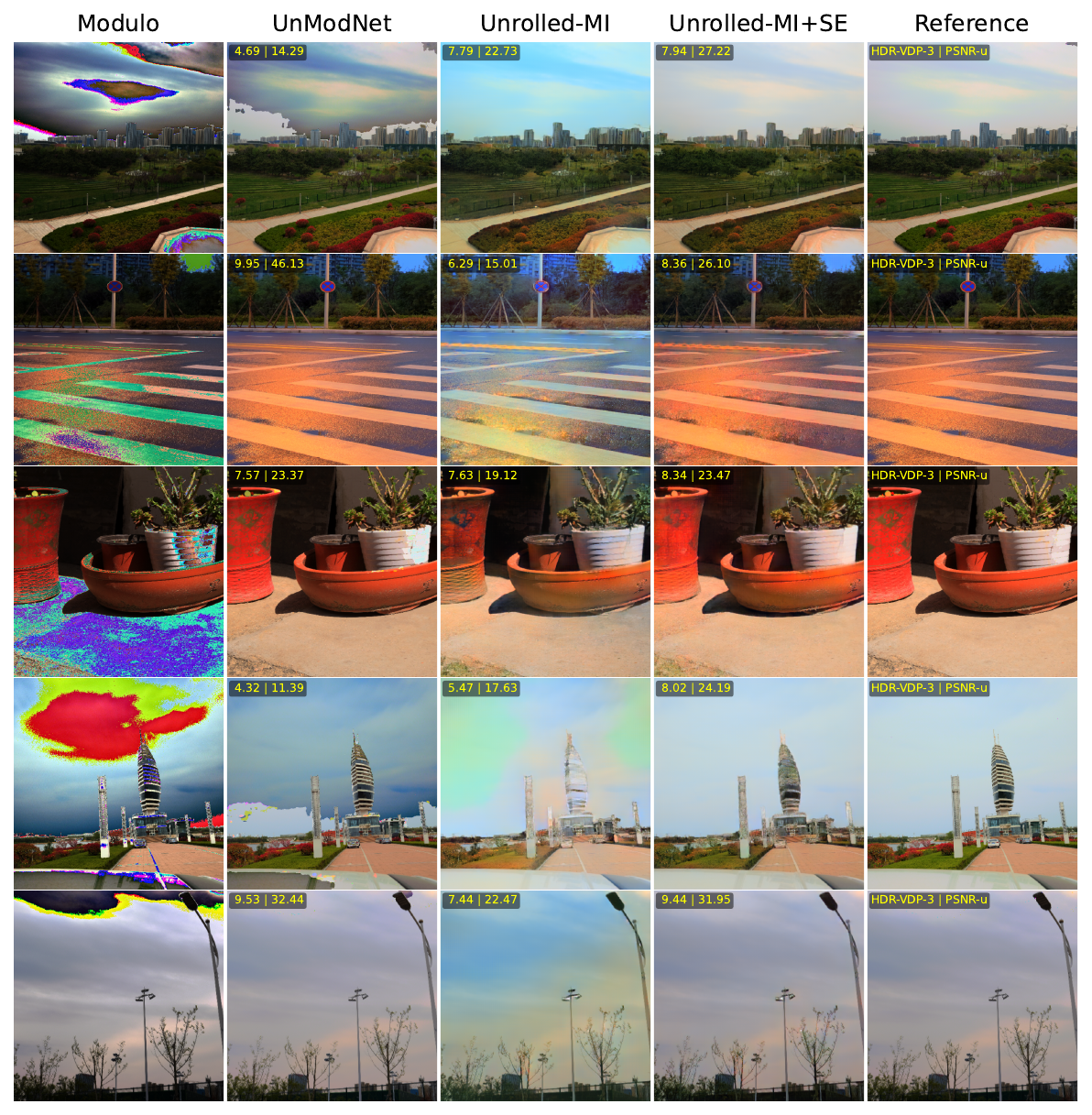} \vspace{-2em}
        \caption{Visual comparison of HDR reconstruction methods on the real RGB scenes from the UnModNet dataset. Columns show input modulo, UnModNet, Unrolled-MI, Unrolled-MI+SE, and the reference. Values denote HDR-VDP3 and PSNR-u metrics per scene.}
        \label{fig:rgb1}
    \end{figure}

\begin{figure}[!h]
        \centering
        \includegraphics[width=\linewidth]{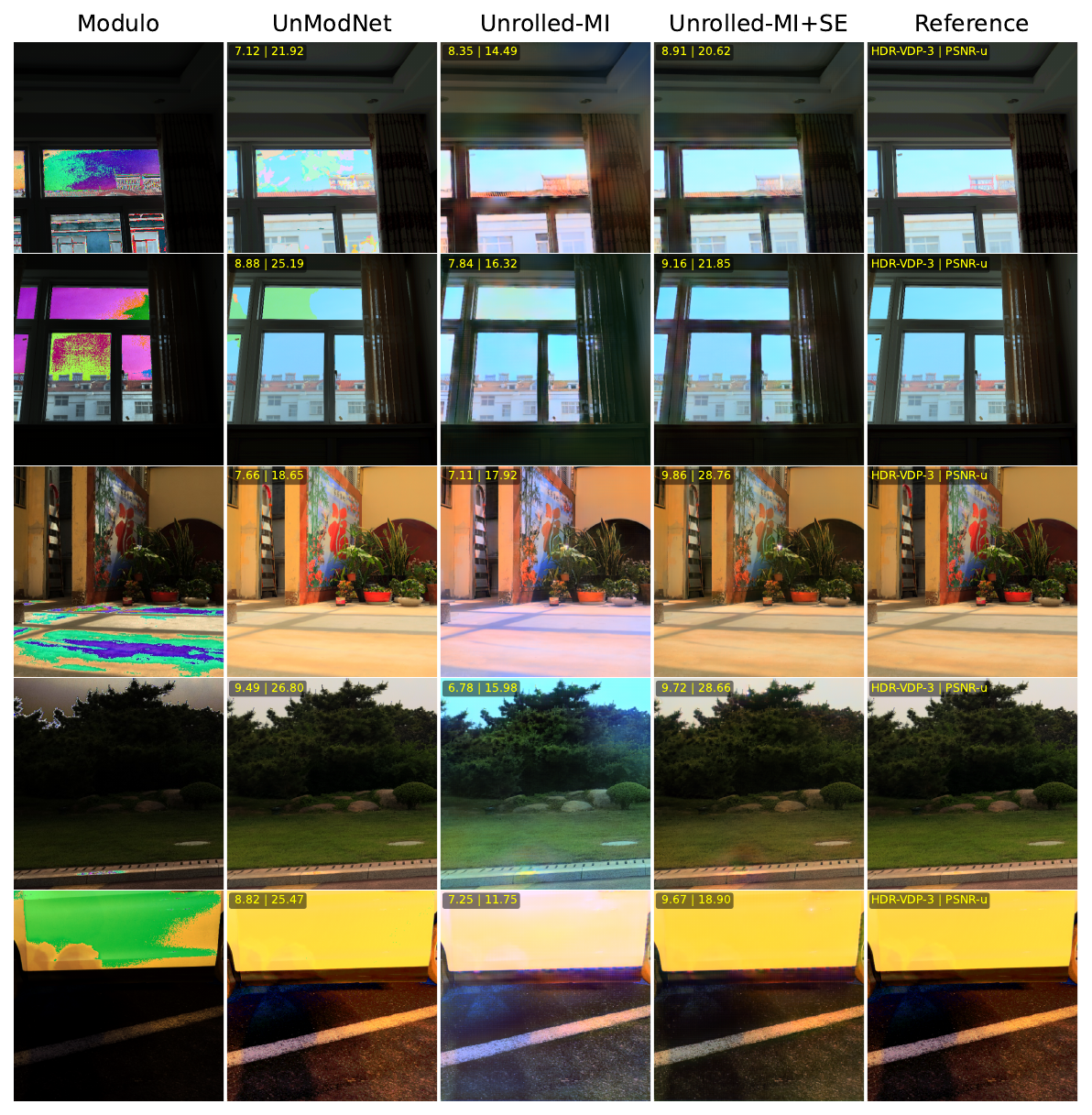}
        \caption{Visual comparison on the ``to improve" real RGB scenes, highlighting challenging cases for all methods. Format and metric annotations are as in Figure 2.}
        \label{fig:rgb2}
    \end{figure}

Figures~\ref{fig:rgb1} and~\ref{fig:rgb2} provide detailed, scene-by-scene comparisons on the full set of real RGB images, highlighting both strengths and areas for improvement. Figure~\ref{fig:rgb1} (``Winning RGB scenes") illustrates scenarios where Unrolled-MI+SE delivers the most accurate visual restoration and highest objective quality, outperforming both UnModNet and the baseline Unrolled-MI. Across diverse indoor and outdoor conditions, Unrolled-MI+SE effectively recovers natural color tones and preserves scene details while substantially reducing typical modulo artifacts and color distortions; this is reflected in the consistently higher HDR-VDP3 and PSNR-u values annotated per image.

\break
Figure~\ref{fig:rgb2} (``To improve RGB scenes") highlights examples where noticeable challenges remain for all algorithms. For certain images with strong luminance transitions or extensive shadows, Unrolled-MI+SE reduces artifacts compared to other methods but can still present residual color shifts or local edge artifacts, especially around window frames and in dark regions. Although Unrolled-MI+SE demonstrates overall improved robustness, these cases reveal the existing limitations in handling deep shadows or very low-light content. Detailed metric values per method accompany each result, supporting a transparent evaluation of both strengths and current limitations. In Figure~\ref{fig:realdata}, we present reconstructions captured using a real modulo camera, encompassing the five real scenes described in the UnModNet paper.

    \newpage
    
\;
\newpage

\begin{figure*}[!t]
    \centering
    \includegraphics[width=\linewidth]{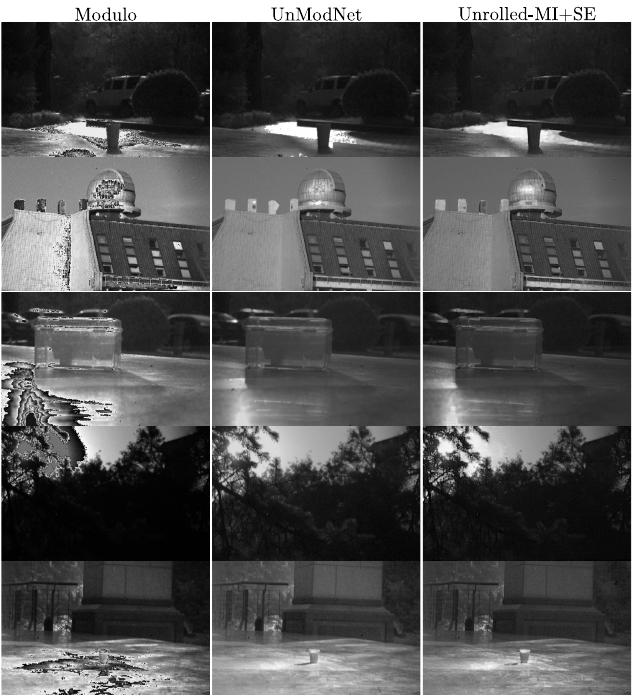}
    \caption{\textbf{Real modulo data evaluation.} Comparison of HDR reconstructions from real modulo sensor captures, contrasting the  performance of UnModNet and the proposed Unrolled-MI+SE method.}
    \label{fig:realdata}
\end{figure*}

\;
\newpage

\noindent 
\begin{minipage}[t]{0.53\textwidth}

\textbf{Different scaling factor $\alpha$ evaluation.} In the Table~\ref{table:3}, we report a quantitative ablation over the modulation factor $\alpha$ and compares our Unrolled-MI + SE model against representative unwrapping and deep learning baselines. Across all values of $\alpha$, our approach consistently attains the highest HDR-VDP-3 scores, with a clear margin over both classical unwrapping (Unwrapped, SPUD, AHFD) and deep learning based (UnModNet). This indicates that the proposed unrolled architecture, combined with the scaling-aware SE regularization, yields reconstructions that are perceptually closer to the reference HDR than competing methods.

In distortion metrics (PSNR-$\mu$, MS-SSIM-$\mu$), UnModNet is a strong baseline. However, Unrolled-MI + SE matches or exceeds UnModNet in PSNR-$\mu$ for almost all $\alpha$, consistently ranks second in MS-SSIM-$\mu$, and achieves higher HDR-VDP-3 indicating a better tradeoff between distortion and perceptual quality (Higher PSNR-$\mu$ for UnModNet at $\alpha=2$ can be attributed to a less challenging unwrapping scenario; specifically, for $\alpha=2$, the signal resets a maximum of only once.
). Varying $\alpha$ from 2.0 to 5.0 only slightly changes rankings: our model maintains the best HDR-VDP-3 and shows minimal MS-SSIM-$\mu$ variation, suggesting robustness to modulation level and that it is not tuned to a narrow regime.
\end{minipage}%
\hfill 
\begin{minipage}[t]{0.45\textwidth} \vspace{-0.8em}
\captionof{table}{Quantitative performance for difference values of $\alpha$. Best and second best results are highlighted in \textbf{bold} and \underline{underline}, respectively.} \centering
\label{table:3}
\resizebox{\linewidth}{!}{
\begin{tabular}{c|l|ccc}
\toprule
$\alpha$  & Method             & HDR-VDP-3 & PSNR-$\mu$ &  MS-SSIM-$\mu$ \\ \midrule  \midrule
2.0 & Unwrapped~\cite{herraez2002fast} & 6.16 & 2.21 & 0.430 \\ 
& SPUD~\cite{pineda2020spud} & 5.87 & 3.42 & 0.545 \\ 
& PnP-UA~\cite{bacca2024deep} & 4.67 & 7.76 & 0.312 \\ 
& AHFD~\cite{monroy2025autoregressive} & 5.36 & 3.53 & 0.582 \\ 
& UnModNet~\cite{zhou2020unmodnet} & \underline{8.26} & \textbf{29.29} & \textbf{0.977} \\ 
& Unrolled-MI + SE   &  \textbf{8.98} & \underline{23.62} &   \underline{0.925} \\ \midrule
3.0 & Unwrapped~\cite{herraez2002fast} & 5.71 & 1.84 & 0.407\\ 
& SPUD~\cite{pineda2020spud} & 5.43 & 3.02 & 0.540 \\ 
& PnP-UA~\cite{bacca2024deep} & 4.73 & 7.72 & 0.305 \\ 
& AHFD~\cite{monroy2025autoregressive} & 4.89 & 2.99 & 0.567 \\ 
& UnModNet~\cite{zhou2020unmodnet} & \underline{7.70} & \underline{21.47} &  \textbf{0.977} \\ 
& Unrolled-MI + SE   &  \textbf{8.79} & \textbf{23.71} &  \underline{0.928} \\ \midrule
4.0 & Unwrapped~\cite{herraez2002fast} & 5.45 & 1.89 & 0.389 \\ 
& SPUD~\cite{pineda2020spud} & 4.97 & 3.00 & 0.539 \\ 
& PnP-UA~\cite{bacca2024deep} & 4.86 & 3.61 & 0.579 \\ 
& AHFD~\cite{monroy2025autoregressive} & 4.60 & 2.48 & 0.532 \\ 
 & UnModNet~\cite{zhou2020unmodnet} & \underline{7.52} & \underline{20.70} & \textbf{0.968} \\ 
& Unrolled-MI + SE   &     \textbf{8.64} & \textbf{22.81} & \underline{0.912} \\ \midrule
5.0 & Unwrapped~\cite{herraez2002fast} & 5.24 & 1.87 & 0.393\\ 
& SPUD~\cite{pineda2020spud} & 4.86 & 2.59 & 0.532 \\ 
& PnP-UA~\cite{bacca2024deep} & 4.60 & 7.58 & 0.292 \\ 
& AHFD~\cite{monroy2025autoregressive} & 4.42 & 2.61 & 0.546 \\ 
& UnModNet~\cite{zhou2020unmodnet} & \underline{7.37} & \underline{20.50} & \textbf{0.960} \\ 
& Unrolled-MI + SE   &  \textbf{8.38} & \textbf{23.53} &  \underline{0.925}  \\ \bottomrule
\end{tabular}
}

\end{minipage}

\section{Discussion and Future Work.}

\textbf{Discussion and Future Work.} While the proposed Unrolled-MI+SE method achieves state-of-the-art performance in HDR modulo imaging, some limitations remain. As demonstrated in the out-of-distribution experiments on real UnModNet RGB scenes, the method shows room for improvement in preserving high-frequency spatial details in dark regions and maintaining accurate color fidelity, particularly in challenging lighting conditions. These artifacts are more pronounced in scenes with abrupt luminance transitions or deep shadows, reflecting intrinsic challenges when recovering higher dynamic range signals compared to standard 8-bit images. Addressing these issues may require future research into color space quantization strategies or integrating learned photographic priors, such as those enabled by methods like NamedCurves~\cite{serrano2024namedcurves}, to enhance tone-mapping and representation quality.

We also observed that incorporating the Scaling Equivariance (SE) regularization when training with in-distribution samples, such as those from the UnModNet dataset, yields only slight improvements in reconstruction metrics. This modest gain is likely due to the dataset's intrinsic diversity in illumination and exposures, which already promotes strong generalization. As such, the primary advantage of SE regularization lies in adapting the model to out-of-distribution data, where it significantly improves generalization and robustness.

Another notable limitation is the need on accurate noise level estimation, which is required as an input parameter for restoration. Overestimation or underestimation of noise can lead to over-smoothed outputs or residual artifacts, respectively. Thus, the robustness of the system in real deployment partly depends on the effectiveness of automatic or adaptive noise estimation. These points indicate delighted directions for future work toward more robust and perceptually faithful HDR reconstruction from modulo measurements.  

\end{document}